\documentclass{article}
\usepackage[utf8]{inputenc}
\usepackage{spconf,amsmath,graphicx}
\usepackage{graphics}
\usepackage{multicol}
\usepackage{lipsum}
\usepackage{color}
\usepackage{amsthm}
\usepackage{amssymb}
\usepackage{mathtools}
\usepackage{amsbsy}
\usepackage{setspace}
\usepackage{float}
\usepackage{lettrine}
\usepackage{newclude}
\usepackage[normalem]{ulem}
\usepackage{latexsym}
\usepackage{multirow}
\usepackage{rotating}
\usepackage{booktabs}
\usepackage{wasysym}
\usepackage{mathrsfs}
\usepackage{bbm} 
\usepackage{epsfig,cite,amsfonts,psfrag}
\usepackage{apptools}
\usepackage{graphicx}
\usepackage[table]{xcolor}
\usepackage{chngcntr}
\usepackage{blindtext}
\usepackage{optidef}
\usepackage{accents}
\usepackage{dsfont}
\usepackage{siunitx}
\usepackage{overpic}
\usepackage{algorithm,algorithmic}

\theoremstyle{plain}

\newtheorem{lemma}{Lemma}

\newtheorem{remark}{Remark}

\newcommand{\argmin}[1]{{\underset{{#1}}{\mathrm{arg\,min}}}}
\newcommand{\vect}[1]{\mathbf{#1}}
\newcommand{\maximize}[1]{{\underset{{#1}}{\mathrm{maximize}}}}
\newcommand{\minimize}[1]{{\underset{{#1}}{\mathrm{minimize}}}}

\def\diag{\mathrm{diag}}

\def\Htran{\mbox{\tiny $\mathrm{H}$}}
\def\Ttran{\mbox{\tiny $\mathrm{T}$}}
\def\CN{\mathcal{N}_{\mathbb{C}}} 

\title{A Novel Discrete Phase Shift Design for RIS-Assisted Multi-User MIMO}

\name{Parisa Ramezani, Yasaman Khorsandmanesh, and Emil Bj{\"o}rnson \thanks{This work was supported by the FFL18-0277 grant from SSF.} }\address{\textit{Department of Computer Science, KTH Royal Institute of Technology, SE-100 44 Stockholm, Sweden}\\ Email: \{parram, yasamank, emilbjo\}@kth.se}

\begin{document}
\maketitle
\ninept

\begin{abstract}
Reconfigurable intelligent surface (RIS) is a newly-emerged technology that might fundamentally change how wireless networks are operated. Though extensively studied in recent years, the practical limitations of RIS are often neglected when assessing the performance of RIS-assisted communication networks. One of these limitations is that each RIS element is restricted to incur a controllable phase shift to the reflected signal from a predefined discrete set. This paper studies an RIS-assisted multi-user multiple-input multiple-output (MIMO) system, where an RIS with discrete phase shifts assists in simultaneous uplink data transmission from multiple user equipments (UEs) to a base station (BS). We aim to maximize the sum rate by optimizing the receive beamforming vectors and RIS phase shift configuration. To this end, we transform the original sum-rate maximization problem into a minimum mean square error (MMSE) minimization problem and employ the block coordinate descent (BCD) technique for iterative optimization of the variables until convergence. We formulate the discrete RIS phase shift optimization problem as a mixed-integer least squares problem and propose a novel method based on sphere decoding (SD) to solve it. Through numerical evaluation, we show that the proposed discrete phase shift design outperforms the conventional nearest point mapping method, which is prevalently used in previous works.
\end{abstract}


\section{Introduction}

The insatiable demand for higher accumulated data rates in wireless communication networks has driven the development of new technological solutions, most recently, reconfigurable intelligent surfaces (RISs). An RIS comprises numerous electrically small reflecting elements, each designed using low-power electronic circuitry such as metamaterials and diodes \cite{Huang2019a,Wu2019a,Renzo2020b}. These elements have the ability to reflect incident signals with adjustable phase shifts, enabling the RIS to manipulate the propagation of incoming waves effectively. The basic use case is to beamforming an incident signal toward the designated receiver \cite{bjornson2022SPM}. However, it can be used to improve the conditions of wireless channels in any given setup. As a result, RISs open a new design dimension in communication engineering and have the potential to revolutionize the operation of wireless networks. 

While RIS can greatly improve network performance in ideal circumstances, it possesses some practical limitations that need to be accounted for when integrating the technology into communication networks. One vital limitation is that practical RIS hardware has a limited bit resolution, meaning that each element can only select phase-shift values from a small discrete set. RIS designs with 1-3 bits are described in \cite{1BITRIS,dai2020reconfigurable,rains2022high}. 
Some recent works have taken this practical constraint into account in their algorithm design and attempted to find desirable discrete phase shifts by solving various optimization problems \cite{Swindlehurst2019,Wu2020b,Di2020,Alexandropoulos2020,HZhang2022}. These problems are inherently hard since discrete phase-shift optimization is a combinatorial problem. Hence, the mentioned works present suboptimal solutions, either by first finding the optimal continuous phase shift configuration and then quantizing each phase to the closest value in the discrete set or by alternatively optimizing one of the phase shifts while keeping the others fixed. To the authors' knowledge, the optimal discrete phase shift configuration is only known in single-user single-antenna systems \cite{Ren2023a}. This paper aims to fill this research gap. 

In this paper, we consider an uplink RIS-assisted multi-user multiple-input multiple-output (MIMO) system where the direct links between the single-antenna user equipments (UEs) and the multi-antenna base station (BS) are obstructed. Hence, the UEs can only communicate with the BS via the RIS. We formulate a sum-rate maximization problem where the objective is to find the optimal receive beamforming vectors and discrete phase shift configuration. Achieving the maximum sum rate has been recognized as a challenging problem with NP-hard complexity \cite{liu2010coordinated}. One widely adopted strategy for addressing this issue is to transform the intractable sum-rate maximization problem into a weighted minimum mean-squared error (MMSE) problem \cite{Shi2011}. We leverage this technique and propose a novel iterative algorithm based on the block coordinate descent (BCD) method for finding a local optimum to the sum-rate maximization problem. Writing the RIS phase shift optimization sub-problem as a mixed integer least squares problem, we propose a novel algorithm based on the sphere decoding (SD) technique to find the optimal discrete RIS phase shifts. As each sub-problem is optimally solved in each iteration of the BCD algorithm, convergence is guaranteed. We show by numerical simulations that the proposed SD-based phase shift optimization outperforms the conventional nearest point mapping method, which is widely adopted in the literature on discrete phase shift design. 

\section{System Model and Problem Formulation}
We study an RIS-assisted multi-user MIMO scenario where a BS equipped with $M$ antennas communicates with $K$ single-antenna UEs through an RIS with $N$ reflecting elements. We consider uplink communication in this paper in which the UEs transmit their data signals to the BS. For brevity, it is also assumed that the direct links between the BS and the UEs are blocked or severely degraded; thus, the communication only takes place via the RIS. All the analysis and results of this paper are extendable to the downlink scenario and to cases where direct links are present. 

We denote the transmitted signal of the $k$th UE (referred to as $\mathrm{U}_k$ henceforth) as $s_k$ and assume an independent and identically
distributed (i.i.d.) complex Gaussian codebook: $s_k \sim CN(0,1)$. The received signal $\vect{y} \in \mathbb{C}^M$ at the BS  can be expressed as \cite{Emil2020RISvsDF}
\begin{equation}
\label{eq:received_signal}
 \vect{y} = \sum_{i=1}^K  \vect{H}\boldsymbol{\Theta}\vect{g}_i \sqrt{p_i}s_i + \vect{n},    
\end{equation}
where $\vect{H} \in \mathbb{C}^{M \times N}$ represents the channel matrix between the BS and the RIS, $\vect{g}_i \in \mathbb{C}^N$ is the channel vector between the RIS and $\mathrm{U}_i$, $p_i$ is the transmit power of $\mathrm{U}_i$, and $\vect{n}\sim \CN(0,\sigma^2 \vect{I}_M)$ is the receiver noise at the BS. Furthermore, $\boldsymbol{\Theta} = \diag (e^{j\theta_1},e^{j\theta_2},\ldots,e^{j\theta_N})$ is the RIS configuration matrix with $\theta_n$ being the phase shift applied to the impinging signal by the $n$th RIS element. The imaginary number is denoted by $j$. Due to hardware limitations, it is not possible for the RIS elements to take any phase values, but the phase shifts can only be picked from a discrete set. We assume that the phase-shift resolution of the RIS is $q$ bits and denote by $\mathcal{D} = \{\hat{\theta}_1,\hat{\theta}_2,\ldots,\hat{\theta}_K\}$ the set of possible phase shifts for each RIS element with $K = 2^q$ different phases. 

To decode the signal from $\mathrm{U}_k$, the BS applies the receive beamforming vector $\vect{a}_k \in \mathbb{C}^M$ to \eqref{eq:received_signal}, which results in
\begin{equation}
 \vect{a}_k^{\Htran} \vect{y} = \vect{a}_k^{\Htran}\vect{H}\boldsymbol{\Theta}\vect{g}_k \sqrt{p_k}s_k + \sum_{i \neq k}  \vect{a}_k^{\Htran}\vect{H}\boldsymbol{\Theta}\vect{g}_i \sqrt{p_i}s_i + \vect{a}_k^{\Htran}\vect{n}.  
\end{equation}
By treating interference as noise, the achievable rate for $\mathrm{U}_k$ can then be obtained as \cite{massivemimobook}
\begin{equation}
    R_k = \log_2(1+\gamma_k), 
\end{equation}
where the signal-to-interference-plus-noise ratio (SINR) is 
\begin{equation}
    \gamma_k = \frac{p_k\left|\vect{a}_k^{\Htran}\vect{H}\diag(\vect{g}_k)\boldsymbol{\theta}\right|^2}{\sum_{i \neq k}p_i \left|\vect{a}_k^{\Htran}\vect{H}\diag(\vect{g}_i)\boldsymbol{\theta}\right|^2 + |\vect{a}_k^{\Htran}\vect{n}|^2}
\end{equation}
and $\boldsymbol{\theta} = \diag (\boldsymbol{\Theta})$ is the RIS phase shift configuration vector.

\subsection{Problem Formulation}
The objective of this paper is to maximize the sum rate of the UEs by jointly optimizing the RIS phase-shift configuration $\boldsymbol{\theta}$ and BS receive beamforming vectors $\vect{A} = [\vect{a}_1,\vect{a}_2,\ldots,\vect{a}_K]$. The problem is formulated as 
\begin{equation}
    \label{eq:main_problem}
    \maximize{\vect{A},\boldsymbol{\theta}}\, \sum_{k=1}^K R_k,~~\textrm{subject to}\,\, \theta_n \in \mathcal{D},\,\forall n,  
\end{equation}
 Problem \eqref{eq:main_problem}
is non-convex due to the multiplication of variables and the discrete phase shift constraint. In this paper, utilizing the well-known equivalence between rate-maximization and MMSE problems, we propose a novel algorithm based on the BCD method that utilizes the SD algorithm to find the optimal discrete phase shifts.
\section{Proposed Solution}
Problem \eqref{eq:main_problem} is non-convex and the global optimum solution is difficult to find. We thus target finding a local optimum by exploiting the well-known equivalence between sum-rate maximization and MSE minimization problems. 
To solve problem \eqref{eq:main_problem}, we first introduce the notation $\vect{G}_k = \vect{H}\diag(\vect{g}_k)$ for the cascaded channel from $\mathrm{U}_k$ to the BS and present the following lemma adapted from \cite{Shi2011} to the problem at hand.
\begin{lemma}
   Let $\omega_k$ be a weight associated with $\mathrm{U}_k$. The sum-rate maximization problem in \eqref{eq:main_problem} is equivalent to the problem 
   \begin{equation}
     \label{eq:equivalent_problem}\minimize{\vect{A},\boldsymbol{\theta},\boldsymbol{\omega}}\, \frac{1}{\ln (2)}\left(\sum_{k=1}^K \omega_k E_k - \ln (\omega_k) \right),~~\mathrm{s.t.}\,\, \theta_n \in \mathcal{D},\,\forall n,    
   \end{equation}
   where $\boldsymbol{\omega} = [\omega_1,\omega_2,\ldots,\omega_K]^{\Ttran}$ and $E_k$ is the mean-squared error (MSE) of $\mathrm{U}_k$'s signal given by 
   \begin{equation}
       \begin{aligned}
       \label{eq:MSE}
         E_k &= \mathbb{E}\left\{|\vect{a}_k^{\Htran}\vect{y} - s_k|^2\right\} = \sum_{i=1}^K p_i|\vect{a}_k^{\Htran}\vect{G}_i\boldsymbol{\theta}|^2 \\ &-\sqrt{p_k}\left( \vect{a}_k^{\Htran}\vect{G}_k \boldsymbol{\theta} + \boldsymbol{\theta}^{\Htran}\vect{G}_k^{\Htran}\vect{a}_k \right) + \sigma^2\|\vect{a}_k\|^2 +1,  
       \end{aligned}
   \end{equation}
   which is obtained by utilizing the mutual independence between $s_1,\dots,s_K$ and between the noise vector $\vect{n}$. 
\end{lemma}
Based on this lemma, we can solve problem \eqref{eq:equivalent_problem}
instead of problem \eqref{eq:main_problem} to find the sum-rate maximizing receive beamforming vectors and RIS phase shift configuration. We solve problem \eqref{eq:equivalent_problem} utilizing the BCD technique. Particularly, given the RIS phase-shift configuration, the problem of finding the optimal receive beamforming vectors reduces to $K$ separate MSE minimization problems as 
\begin{equation}
  \minimize{\vect{a}_k}\, E_k,~~k = 1,2,\ldots,K,   
\end{equation}for which the solution is the classical MMSE receive beamforming vector \cite{massivemimobook}:
\begin{equation}
\label{eq:optimal_receive_beamformer}
   \vect{a}_k^* = \sqrt{p_k}\left( \sum_{i=1}^K p_i \vect{G}_i \boldsymbol{\theta}\boldsymbol{\theta}^{\Htran}\vect{G}_i^{\Htran} + \sigma^2 \vect{I}_M\right)^{-1} \vect{G}_k \boldsymbol{\theta}. 
\end{equation}
Next, for given receive beamforming vectors and RIS phase-shift configuration, the optimal value for $\omega_k$ is obtained as 
\begin{equation}
\label{eq:optimal_weight}
    \omega_k^* = \frac{1}{E_k}.
\end{equation}
Finally, when the receive beamforming vectors $\vect{a}_k$ and the weights $\omega_k$ are given for $k=1,\ldots,K$, the problem of optimizing the RIS phase-shift configuration is expressed as 
\begin{equation}
\label{eq:phase_optimization_problem}
\begin{aligned}
 &\minimize{\boldsymbol{\theta}}\quad\boldsymbol{\theta}^{\Htran}\vect{B}\boldsymbol{\theta} - \boldsymbol{\theta}^{\Htran}\vect{b} - \vect{b}^{\Htran}\boldsymbol{\theta},\\
 &\textrm{subject to}\,\quad \theta_n \in \mathcal{D},\,\forall n, 
 \end{aligned}
\end{equation}
where
\begin{equation}
    \begin{aligned}
    \label{eq:B_and_b}
     \vect{B} &= \sum_{k=1}^K \omega_k \sum_{i=1}^K p_i \vect{G}_i^{\Htran}\vect{a}_k \vect{a}_k^{\Htran} \vect{G}_i,\\
     \vect{b}& = \sum_{k=1}^K \omega_k \sqrt{p_k}\vect{G}_k^{\Htran}\vect{a}_k. 
    \end{aligned}
\end{equation}
Although the objective function has a quadratic form, problem \eqref{eq:phase_optimization_problem} is not a convex optimization problem due to the discrete constraint on the RIS phase shifts. We will first present a heuristic solution to \eqref{eq:phase_optimization_problem} and then our main proposed algorithm.

\subsection{Heuristic Solution}
If we drop the discrete constraint in \eqref{eq:phase_optimization_problem}, the solution to the relaxed optimization problem is  
\begin{equation}
\label{eq:theta_hat}
    \hat{\boldsymbol{\theta}} = \vect{B}^{\dagger}\vect{b},
\end{equation}
where $\vect{B}^{\dagger}$ denotes the pseudo-inverse of $\vect{B}$. We need the pseudo-inverse since $\vect{B}$ might be rank-deficient, although $\vect{b}$ is in its span.
 One heuristic way to solve problem \eqref{eq:phase_optimization_problem} is to drop the amplitude of the entries of $\hat{\boldsymbol{\theta}}^*$ and then map each of them to its nearest feasible point in the set $ e^{j\mathcal{D}}$, similar to \cite{Alexandropoulos2020}. This heuristic approach produces the sub-optimal discrete RIS phase configuration vector, obtained as

\begin{equation}
    \label{eq:nearest_point}
    \theta^*_n = \argmin{\theta_n \in \mathcal{D}}\, \Big|e^{j\arg([\hat{\boldsymbol{\theta}}]_n)} - e^{j\theta_n}\Big|,~~n=1,2,\ldots,N.
\end{equation}
This simple approach has resulted in promising results in previous works, but it is sub-optimal due to the quantization of a hypothetical continuous RIS configuration and because the discrete phases are selected independently so that the quantization errors are piling up. 

\subsection{Proposed SD-Based Solution}
Herein, we propose a novel solution to \eqref{eq:phase_optimization_problem} based on the SD algorithm from MIMO detection theory \cite{jalden2005complexity} to directly find the optimal discrete phase shift values instead of quantizing a continuous vector. To this end, we re-write the objective function of \eqref{eq:phase_optimization_problem} as 

\begin{equation}
    \|\vect{c} - \vect{R}\boldsymbol{\theta}\|^2 - \vect{c}^{\Htran}\vect{c},
\end{equation}where $\vect{R}\in \mathbb{C}^N$ is an upper triangular matrix obtained from the Cholesky decomposition of $\vect{B}$ as $\vect{B} = \vect{R}^{\Htran}\vect{R}$ and $\vect{c} = (\vect{b}^{\Htran}\vect{R}^{-1})^{\Htran}$. Using this notation, the RIS phase shift optimization problem in \eqref{eq:phase_optimization_problem} can be recast as 
\begin{equation}
\label{eq:SD_problem}
\minimize{\boldsymbol{\theta}}
\, \|\vect{c} - \vect{R}\boldsymbol{\theta}\|^2,~~\textrm{subject to}\,\, \theta_n \in \mathcal{D},\,\forall n.
\end{equation}
The problem \eqref{eq:SD_problem} is classified as an integer least-squares problem due to the discrete phase shifts. The search space is a finite subset of the set of infinite-resolution RIS phase shifts. To solve this problem efficiently, we utilize the SD technique which has been proposed as a technique for solving closest lattice point problems with lower computational complexity compared to a naive exhaustive search \cite{jalden2005complexity}. SD achieves this lower complexity by reducing the number of search points within the skewed lattice within a hypersphere of radius $d$, ensuring an optimal solution is still obtained. In \eqref{eq:SD_problem},  the triangular structure of $\vect{R}$ allows us to utilize SD. 

In this paper, we propose to use the Schnorr-Euchner SD (SESD) algorithm \cite{Agrell2002}, which employs a zig-zag enumeration approach to sort candidate points. SESD improves upon the basic SD algorithm by initially examining the smallest child node of each parent node in each layer. This strategy exploits the fact that the first feasible solution found is often highly suitable, enabling a rapid reduction in the search radius. Consequently, numerous branches can be pruned, resulting in a further reduction of computational complexity. 

When the matrix $\vect{B}$ is rank-deficient, zero-valued elements appear on the diagonal of the upper-triangular matrix $\vect{R}$ and standard SD algorithms are not applicable for solving \eqref{eq:SD_problem}. Several generalized SD (GSD) algorithms have been proposed to deal with such under-determined problems.  In this paper, we use the approach proposed in \cite{cui2004efficient} and add a constant term $\alpha N$ to the objective function in \eqref{eq:phase_optimization_problem} to obtain an equivalent problem with a full-rank upper-triangular matrix $\hat{\vect{R}}$. Specifically, the problem in \eqref{eq:phase_optimization_problem} is equivalent to 
\begin{equation}
\label{eq:phase_optimization_problem2}
\begin{aligned}
&\minimize{\boldsymbol{\theta}}\quad\boldsymbol{\theta}^{\Htran}\vect{B}\boldsymbol{\theta} - \boldsymbol{\theta}^{\Htran}\vect{b} - \vect{b}^{\Htran}\boldsymbol{\theta} + \alpha \|\boldsymbol{\theta}\|^2,\\
 &\textrm{subject to}\,\quad \theta_n \in \mathcal{D},\,\forall n.
 \end{aligned}
\end{equation}$\alpha \|\boldsymbol{\theta}\|^2$ is a constant since $\|\boldsymbol{\theta}\|^2 = N$. Therefore, solving problem \eqref{eq:phase_optimization_problem2} is equivalent to solving $\eqref{eq:phase_optimization_problem}$ since they only differ in a constant term. Now, we can re-write problem $\eqref{eq:phase_optimization_problem2}$ as an integer least-squares problem similar to the one in \eqref{eq:SD_problem}:
\begin{equation}
\label{eq:GSD_problem}
\minimize{\boldsymbol{\theta}}
\, \|\hat{\vect{c}} - \hat{\vect{R}}\boldsymbol{\theta}\|^2,~~\textrm{subject to}\,\, \theta_n \in \mathcal{D},\,\forall n,
\end{equation}where $\vect{B} + \alpha \vect{I}_N = \hat{\vect{R}}^{\Htran} \hat{\vect{R}}$ and $\hat{\vect{c}} = (\vect{b}^{\Htran}\hat{\vect{R}}^{-1})^{\Htran}$. As the diagonal elements in $\hat{\vect{R}}$ are non-zero, standard SD algorithms can be applied to solve \eqref{eq:GSD_problem}.

\begin{remark}
The choice of the parameter $\alpha$ influences the complexity and there exists an optimal $\alpha$ value that minimizes the complexity. However, it is not straightforward to derive a closed-form expression for the optimal $\alpha$ \cite{cui2004efficient}. In our simulations, we have set $\alpha = 1$.
\end{remark}

Using the procedure described above, we alternately optimize the receive beamforming matrix $\vect{A}$, the weights $\boldsymbol{\omega}$, and the RIS phase shift configuration vector $\boldsymbol{\theta}$ until a satisfactory convergence is achieved. Algorithm \ref{Alg:BCD} summarizes the proposed procedure.

\begin{algorithm}[t!]
\small
\caption{BCD algorithm for solving \eqref{eq:equivalent_problem}.}
\label{Alg:BCD}
\begin{algorithmic}[1]
\STATE{Initialize RIS phase shift configuration $\boldsymbol{\theta}^{(0)}$.}
\STATE{Set the solution accuracy $\epsilon>0$.}
\STATE{Set $\Delta = \epsilon + 1$, $l =0$.}
\WHILE{$\Delta > \epsilon$}
\STATE{$l=l+1$.}
\STATE{Obtain the receive beamforming vectors $\{\vect{a}_k^{*(l)}\}$ from \eqref{eq:optimal_receive_beamformer}.}
\STATE{Update the MSE values based on $\{\vect{a}_k^{*(l)}\}$ and obtain the optimal weights $\{\omega_k^{*(l)}\}$ from \eqref{eq:optimal_weight}.}
\STATE{Substitute $\{\vect{a}_k^{*(l)}\}$ and $\{\omega_k^{*(l)}\}$ into \eqref{eq:B_and_b}, find the upper triangular matrix $\vect{R}$ (or $\hat{\vect{R}}$) and the vector $\vect{c}$ (or $\hat{\vect{c}}$), and obtain $\boldsymbol{\theta}^{*(l)}$ by solving problem \eqref{eq:SD_problem} (or \eqref{eq:GSD_problem}) using the SESD technique.}
\STATE{Compute $\{E_k^{(l)}\}$ by substituting $\vect{a}_k^{*(l)}$ and $\boldsymbol{\theta}^{*(l)}$ into \eqref{eq:MSE}.}
\STATE{Compute $f^{(l)} = \sum_{k=1}^K \omega_k^{*(l)}E_k^{(l)} - \ln (\omega_k^{*(l)})$}.
\STATE{$\Delta = |f^{(l)} - f^{(l-1)}|$.}
\ENDWHILE
\RETURN $\vect{a}_k^{\mathrm{opt}} = \vect{a}_k^{*(l)}~\forall k$, $\boldsymbol{\theta}^{\mathrm{opt}} = \boldsymbol{\theta}^{*(l)}$.

 \end{algorithmic}
\end{algorithm}
\begin{remark}
    The BCD algorithm can guarantee convergence to a stationary point when the optimal solution to each sub-problem is obtained in each iteration \cite{Tseng2001}. This is the case with our proposed algorithm as the three sub-problems for finding the BS receive beamforming vectors, associated weights, and RIS phase shift configuration are all optimally solved.   
\end{remark}
\section{Numerical Results}
In this section, we evaluate the performance of the proposed BCD approach presented in Algorithm~\ref{Alg:BCD}. The heuristic nearest point method is used as a benchmark, where RIS phase shifts in step 8 of Algorithm~\ref{Alg:BCD} are obtained from \eqref{eq:nearest_point}.
\subsection{Simulation Setup}
An RIS-assisted multi-user MIMO system is considered with $K=6$ UEs. Unless otherwise stated, the following setup is used throughout the simulations. The BS has a uniform linear array configuration and is equipped with $M=10$ antennas, while the RIS has $N = 64$ or $N=32$ reflecting elements, with $8$ elements in the horizontal orientation and either $8$ or $4$ elements in the vertical orientation. The channels are modeled by Rician fading. Specifically, the channel between the RIS and the BS is given by 
\begin{equation}
\label{eq:Rician_fading_channel}
    \vect{H} = \sqrt{\beta_{H}} \left(\sqrt{\frac{\kappa_H}{\kappa_H + 1}} \vect{H}_{\mathrm{LOS}} + \sqrt{\frac{1}{\kappa_H + 1}} \vect{H}_{\mathrm{NLOS}}\right),
\end{equation}
where $\kappa_H$ is the Rician factor set as $\kappa_H = 5$, and $\vect{H}_{\mathrm{LOS}}$ and $\vect{H}_{\mathrm{NLOS}}$ are the LOS and NLOS components of $\vect{H}$. In particular, $\vect{H}_{\mathrm{LOS}} = \vect{r}_{\mathrm{BS}}(\vartheta)\vect{r}_{\mathrm{RIS}}^{\Ttran}(\varphi_{\mathrm{AOD}},\phi_{\mathrm{AoD}})$ where $\vect{r}_{\mathrm{BS}}$ and $\vect{r}_{\mathrm{RIS}}$ respectively represent the array response vectors of the BS and the RIS, with $\vartheta$ being the angle of arrival (AoA) at the BS, $\varphi_{\mathrm{AoD}}$ and $\phi_{\mathrm{AoD}}$ being the azimuth and elevation angle of departure (AoD) from the RIS \cite[Chapter 7]{massivemimobook}. The angles are set as $\vartheta = \pi/6$, $\varphi_{\mathrm{AoD}} = -\pi/3$, and $\phi_{\mathrm{AoD}} = \pi/6$. The spacing between the RIS elements  and the BS antennas are set as $\lambda/4$ and $\lambda/2$, respectively, where $\lambda$ is the wavelength. For $\vect{H}_{\mathrm{NLOS}}$, we consider correlated Rayleigh fading and use the local scattering spatial correlation model with Gaussian distribution \cite[Chapter 2]{massivemimobook}, where the scattering is assumed to be around the RIS and in the azimuth domain. Furthermore, $\beta_{H}$ denotes the path-loss and is modeled as \cite{Emil2020RISvsDF}
\begin{equation}
     \beta_H = -37.5 - 22 \log_{10} \big(d_H/1~ \mathrm{m}\big) ~~[\mathrm{dB}],
\end{equation} with the carrier frequency of $3\,$ GHz. $d_H$ is the distance between the BS and the RIS, which is set as $d_H=20\,$m. The channels between the UEs and the RIS are modeled in a similar way with $\kappa_g $, $\varphi_{\mathrm{AoA},k}$, $\phi_{\mathrm{AoA},k}$, and $d_{g,k}$ being the Rician factor, azimuth AoA from $\mathrm{U}_k$ to the RIS, elevation AoA from $\mathrm{U}_k$ to the RIS, and the distance between $\mathrm{U}_k$ and the RIS. We set $\kappa_g = 5$ and assume that the UEs are uniformly distributed around the RIS such that $d_{g,k} \sim \mathcal{U}[20 ~\mathrm{m},40 ~\mathrm{m}]$, $\varphi_{\mathrm{AoA},k} \sim \mathcal{U}[-\pi/3,\pi/3]$, and $\phi_{\mathrm{AoA},k} \sim \mathcal{U}[-\pi/6,0]$. The noise power is $\sigma^2 = -160\,$dBm. Furthermore, the convergence threshold is set as $\epsilon = 10^{-3}$ and $\boldsymbol{\theta}^{(0)}$ in Algorithm \ref{Alg:BCD} is initialized in a random fashion such that the phase shift of each element is randomly picked from the set of discrete phase shifts. The discrete set of uniformly spaced RIS phase shifts is given by 
\begin{equation}
   \mathcal{D} = \left\{\frac{m\pi}{2^{q-1}}: m =0,1,\ldots,2^q - 1 \right\}.
\end{equation}We set the bit resolution of RIS elements as $q = 1$; thus, we have $\mathcal{D} = \{0,\pi\}$.

\begin{figure}[t!]
    \centering
    \includegraphics[width= 0.9\linewidth]{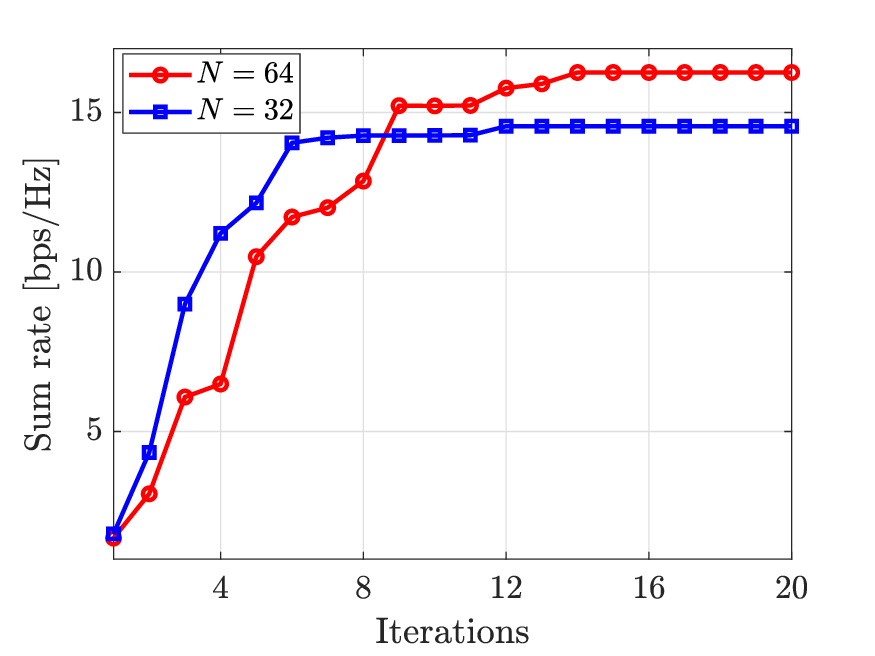} 
    \caption{Sum rate evolution over iterations for different number of RIS elements.}
    \label{fig:convergence}
\end{figure}

\begin{figure}[t!]
    \centering
    \includegraphics[width= 0.9\linewidth]{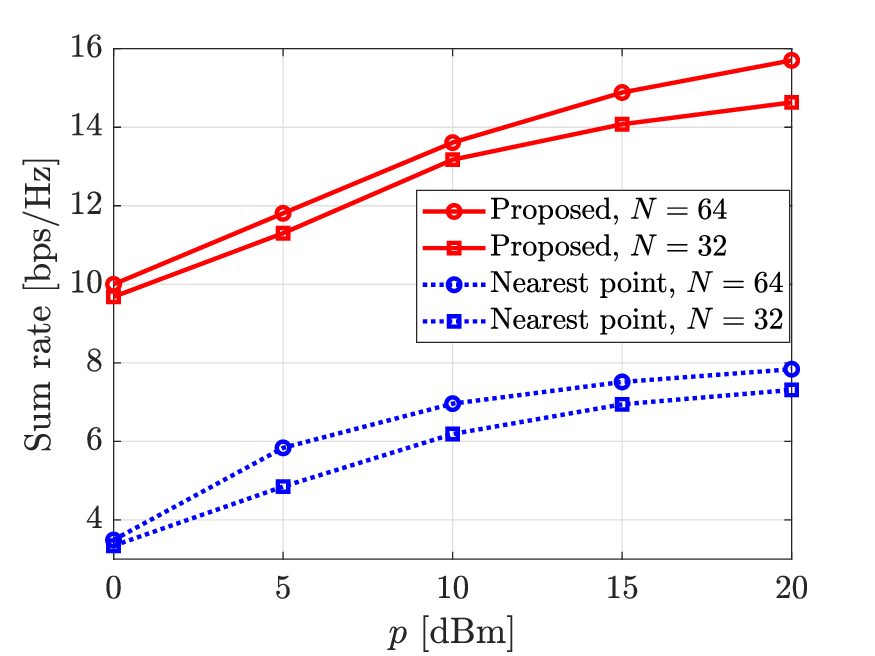} 
    \caption{Sum rate versus the UEs' transmit power.}
    \label{fig:SESD_vs_Nearest}
\end{figure}

\subsection{Results and Discussion}
Fig.~\ref{fig:convergence} presents the convergence behavior of Algorithm \ref{Alg:BCD} for two different number of RIS elements: $N = 64$ and $N = 32$. The UEs' transmit power is given by $p_i = 20$\,dBm, $,~\forall i$. The proposed BCD algorithm reaches a stationary point after $12$ iterations when $N=32$ and $14$ iterations in case of $N=64$. When $N=64$, the sum rate converges to a higher value which is due to the increased aperture gain and beamforming gain as well as better interference cancellation capability of a RIS with greater number of elements. 

Fig.~\ref{fig:SESD_vs_Nearest} evaluates the sum rate performance versus the UEs' transmit power where $p_i = p,~\forall i$. 
We compare the proposed SD-based discrete phase shift optimization method with the heuristic nearest point method which finds the discrete RIS phase shifts by quantizing the optimized continuous ones. It is evident from Fig.~\ref{fig:SESD_vs_Nearest} that the proposed SD-based phase shift solution outperforms the nearest point-based design because the former finds the optimal RIS discrete phase shifts in each iteration of Algorithm \ref{Alg:BCD}, while the latter can only obtain a sub-optimal solution by quantizing the optimized continuous phases. In fact, with the RIS phase shifts obtained via the nearest point method, the interference among the users cannot be properly resolved, resulting in very low data rates.

\section{Concluding Remarks}
In this paper, we proposed a novel discrete phase-shift optimization method for RIS-assisted multi-user MIMO systems. By transforming the original sum-rate maximization problem into an equivalent MMSE minimization problem, we presented a BCD algorithm to iteratively optimize the BS receive beamforming vectors and the RIS phase shift configuration. Particularly, for the RIS phase shift optimization, we proposed a novel method based on the SESD algorithm where the optimal discrete RIS phase shifts can be obtained in each iteration. We have numerically shown that the proposed strategy boosts the network performance compared to using the conventional nearest point mapping for finding the sub-optimal discrete RIS phase shifts. This highlights the importance of jointly optimizing the discrete phase shifts instead of separate quantization of the entries of a continuous vector. The proposed methodology can be applied to solve many similar discrete phase-shift design problems.

\bibliographystyle{IEEEtran}
\bibliography{refs}
\end{document}